\documentclass[12pt]{article}
\usepackage{graphicx}
\usepackage{url}
\usepackage{xspace}
\usepackage{defs}

\newcommand\pubnumber{}
\newcommand\pubdate{\today}

\def\institute{Regional Centre of Advanced Technologies and Materials, Joint Laboratory of
Optics of Palack\'{y} University and Institute of Physics AS CR, Faculty of Science,
Palack\'{y} University, 17. listopadu 12, 771 46 Olomouc, Czech Republic}
\def\support{\footnote{On behalf of the ATLAS collaboration, copyright 2020 CERN for the benefit of the ATLAS Collaboration.
CC-BY-4.0 license.}}

\def\Title#1{\begin{center} {\Large #1 } \end{center}}
\def\Author#1{\begin{center}{ \sc #1} \end{center}}
\def\Address#1{\begin{center}{ \it #1} \end{center}}

\newcommand\pubblock{\rightline{\begin{tabular}{l} \pubnumber\\
         \pubdate  \end{tabular}}}
\newenvironment{Abstract}{\begin{quotation}  }{\end{quotation}}
\newenvironment{Presented}{\begin{quotation} \begin{center} 
             PRESENTED AT\end{center}\bigskip 
      \begin{center}\begin{large}}{\end{large}\end{center} \end{quotation}}
\def\Acknowledgements{\bigskip  \bigskip \begin{center} \begin{large}
             \bf ACKNOWLEDGEMENTS \end{large}\end{center}}

\def\beq{\begin{equation}}
\def\eeq#1{\label{#1}\end{equation}}
\def\eeqn{\end{equation}}

\def\beqa{\begin{eqnarray}}
\def\eeqa#1{\label{#1}\end{eqnarray}}
\def\eeqan{\end{eqnarray}}

\let\bar=\overbar

\def\Dslash{\not{\hbox{\kern-4pt $D$}}}
\def\dslash{\not{\hbox{\kern-2pt $\del$}}}

\def\msb{{\bar{\ssstyle M \kern -1pt S}}}

\begin{document}
\begin{titlepage}
\pubblock

\vfill
\Title{Single- and and double-differential cross-sections in $t\bar{t}$ final states in the $\ell$+jets channel}
\vfill
\Author{  Ji\v{r}\'{\i} Kvita\support}
\Address{\institute}
\vfill
\begin{Abstract}
We present measurements of single- and double-differential cross-sections in $\ttbar$ final states in the lepton+jets channel, unfolded to particle and parton levels with both resolved and boosted topologies considered. Fully corrected spectra are presented as functions of the top-quark as well as $\ttbar$ system kinematic variables and jet multiplicities in data from pp collisions at $\sqrt{s}=13\,$TeV collected in 2015 and 2016 by the ATLAS detector at the LHC, corresponding to integrated luminosity of $36\,\mathrm{fb}^{-1}$ . Measurements provided detailed information on the top-quark production and decay, enabling precision tests of modern Monte Carlo generators as well as latest fixed-order Standard Model predictions. Good agreement between the theoretical predictions and the data is observed within reduced systematic uncertainties w.r.t. previous ATLAS measurements.
\end{Abstract}
\vfill
\begin{Presented}
$12^\mathrm{th}$ International Workshop on Top Quark Physics\\
Beijing, China, 22--27 September 2019.
\end{Presented}
\vfill
\end{titlepage}
\def\thefootnote{\fnsymbol{footnote}}
\setcounter{footnote}{0}
%

\section{Introduction}

  We present measurements~\cite{Aad:2019ntk} of single- and double-differential cross-sections in $t\bar{t}$ final states in the $\ell$+jets channel, corrected for detector resolution and acceptance effects, with both resolved and boosted topologies considered.
  Fully corrected spectra are presented as functions of the top-quark as well as $t\bar{t}$ system kinematic variables and jet multiplicities in data from $pp$ collisions at $\sqrt{s}=13\,\mathrm{TeV}$ collected in 2015 and 2016 by the ATLAS detector~\cite{Aad:2008zzm} at the LHC, corresponding to an integrated luminosity of $36\,\mathrm{fb}^{-1}$.

  \section{Data and simulation}
 The \Powheg+\PythiaEight generator is used for the simulation of the $t\bar{t}$ events to model corrections which are applied to data. Data-driven techniques are used to model the multijet background while simulation is used for modelling of the $W/Z+$jets and single top processes. A total of 1.3M (48k) events in the resolved (boosted) regime is observed in the data, with expected purities of 89\% (86\%), leading to a~1\% (8\%) data/prediction agreement. Worse agreement in the boosted regime is due to the long standing observation of the softer top quark transverse momentum spectrum in the data compared to the NLO predictions.

\section{Event reconstruction}
In the resolved topology, the pseudotop algorithm~\cite{Aad:2015eia} is applied to form proxy pseudotop objects from jets, leptons and missing transverse energy, with a high correlation to the kinematics of the original top quarks, but allowing the study of the top quark at the particle level. 
In the boosted regime, the reconstructed high-$p_\mathrm{T}$ large-$R$ jet (clustered using small-$R$ jets as constituents) serves as the hadronic top quark candidate, while the leptonic top quark four-momentum is reconstructed in the same way as in the resolved case by adding the momenta of the $b$-jet angularly closer to the lepton and leptonically-decaying $W$ boson, where the neutrino longitudinal momentum is computed from the $m_{\ell\nu} = m_W$ condition.

\section{Unfolding}
In order to correct the detector-level observed spectra for finite detector acceptance, efficiency and resolution, an unfolding procedure is applied.
 Both resolved and boosted topologies are measured, with both particle and parton level results obtained. Also, both absolute as well as relative (normalized) cross-sections are measured.
The unfolding procedure can be summarized as follows
$$ \frac{\mathrm{d}^2\sigma_i}{\mathrm{d}X\,\mathrm{d}Y} = \frac{1}{\mathcal{L} \Delta X \Delta Y}\frac{1}{\epsilon_i} \mathcal{M}^{-1}_{ij}\, f_j^\mathrm{acc} \,(D-B)_j \,,$$
where $\mathcal{M}^{-1}$ stands not for a direct matrix inversion but rather for a regularized matrix inversion (unfolding) using the iterative Bayesian method. The efficiency $\epsilon_i$ and acceptance $f_j^\mathrm{acc}$ correct for events escaping the fiducial phase space in particle-level bin $i$ and detector-level bin $j$, respectively.
In the resolved topology, an additional matching correction ensures that objects forming the pseudo-top quarks are angularly well matched between the particle and detector levels. In unfolding to the parton level, the acceptance stands for dilepton events removal. In the resolved topology, parton level equals the full phase space, reconstructed using the KLFitter~\cite{KLFit:2013}; in the boosted regime, the phase space is limited to partonic top quark $p_\mathrm{T} > 350\,$GeV.

\section{Results}
\begin{figure}
  \includegraphics[width=0.245\linewidth]{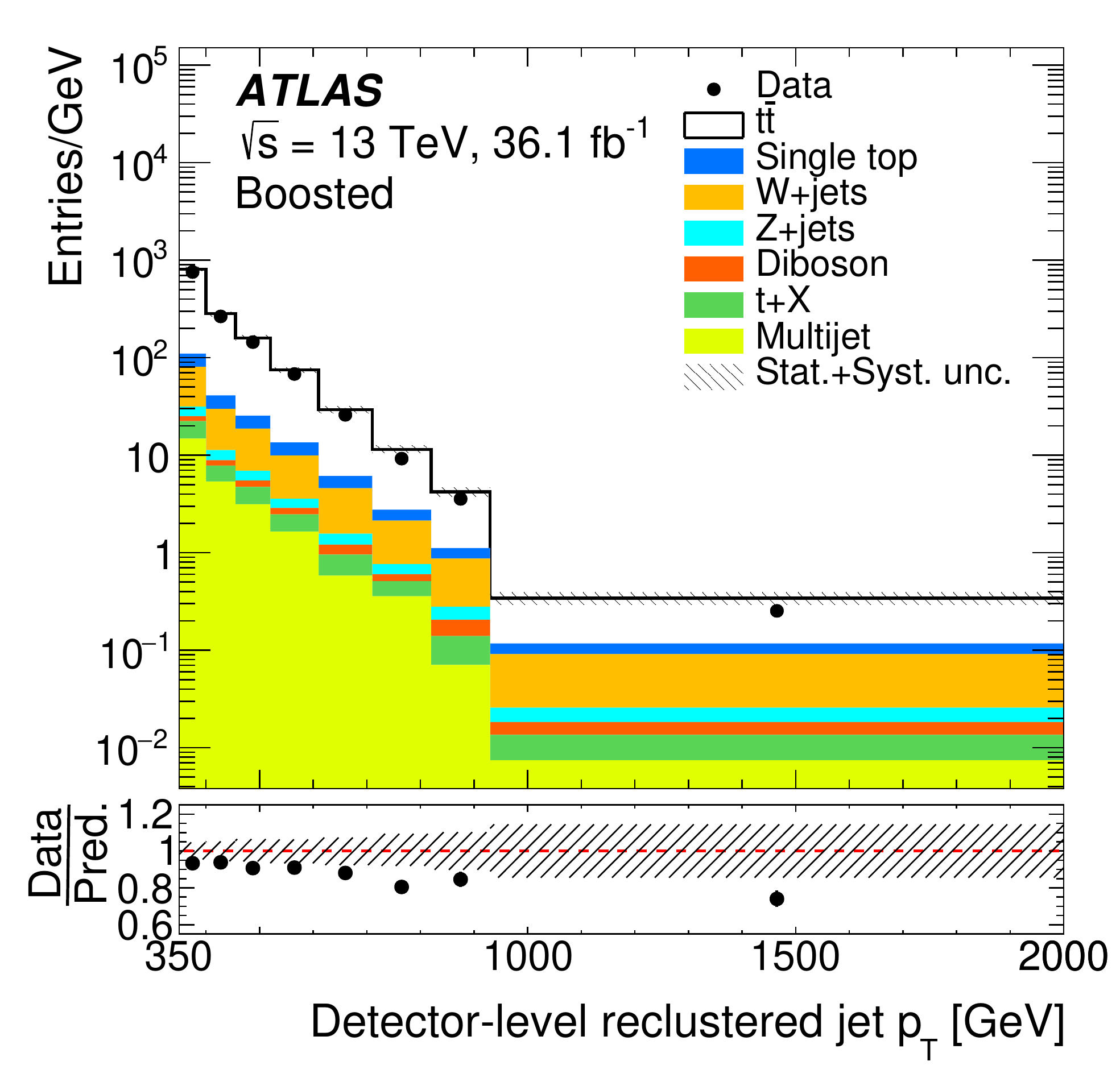}
  \includegraphics[width=0.245\linewidth]{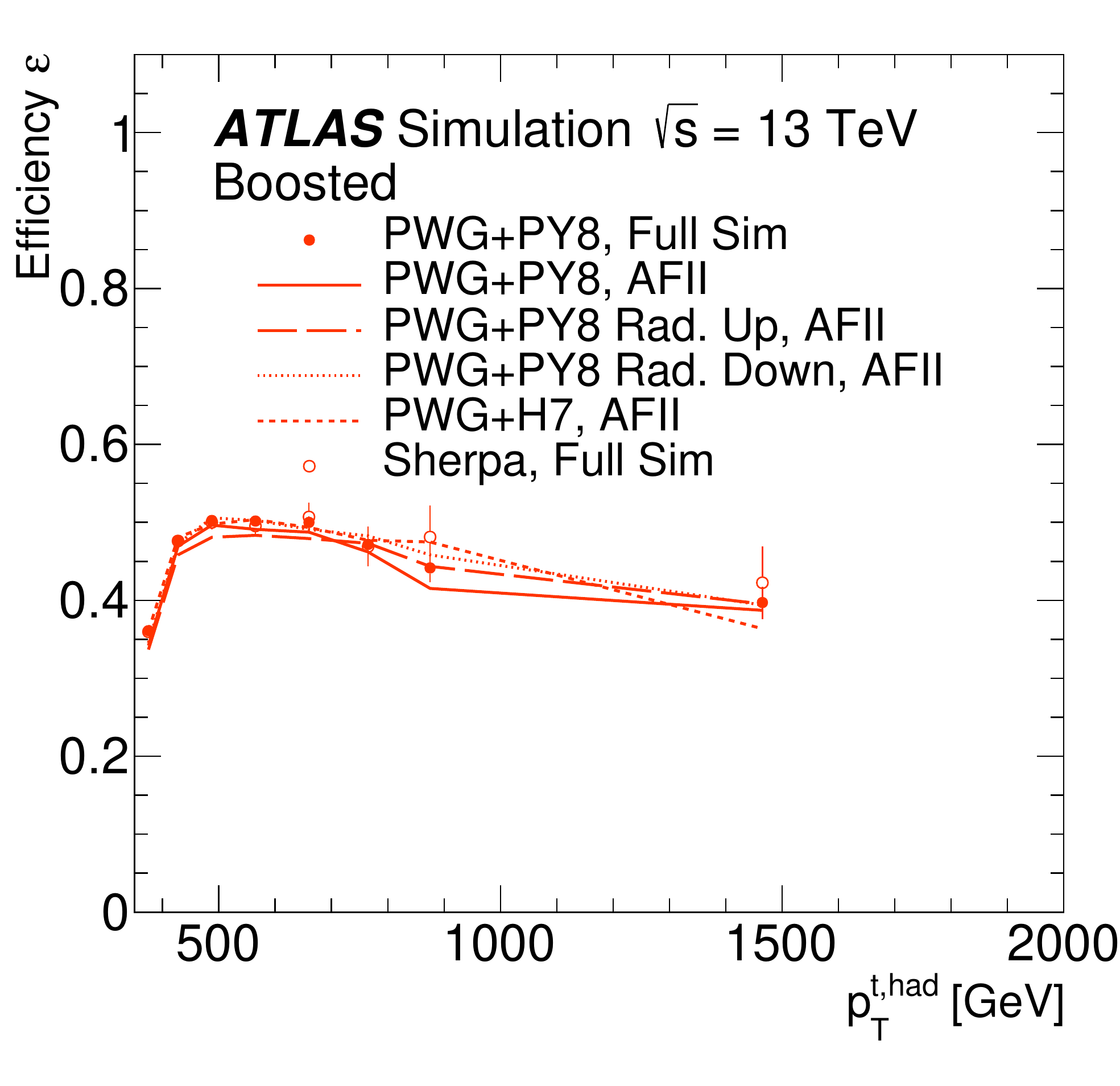}
  \includegraphics[width=0.245\linewidth]{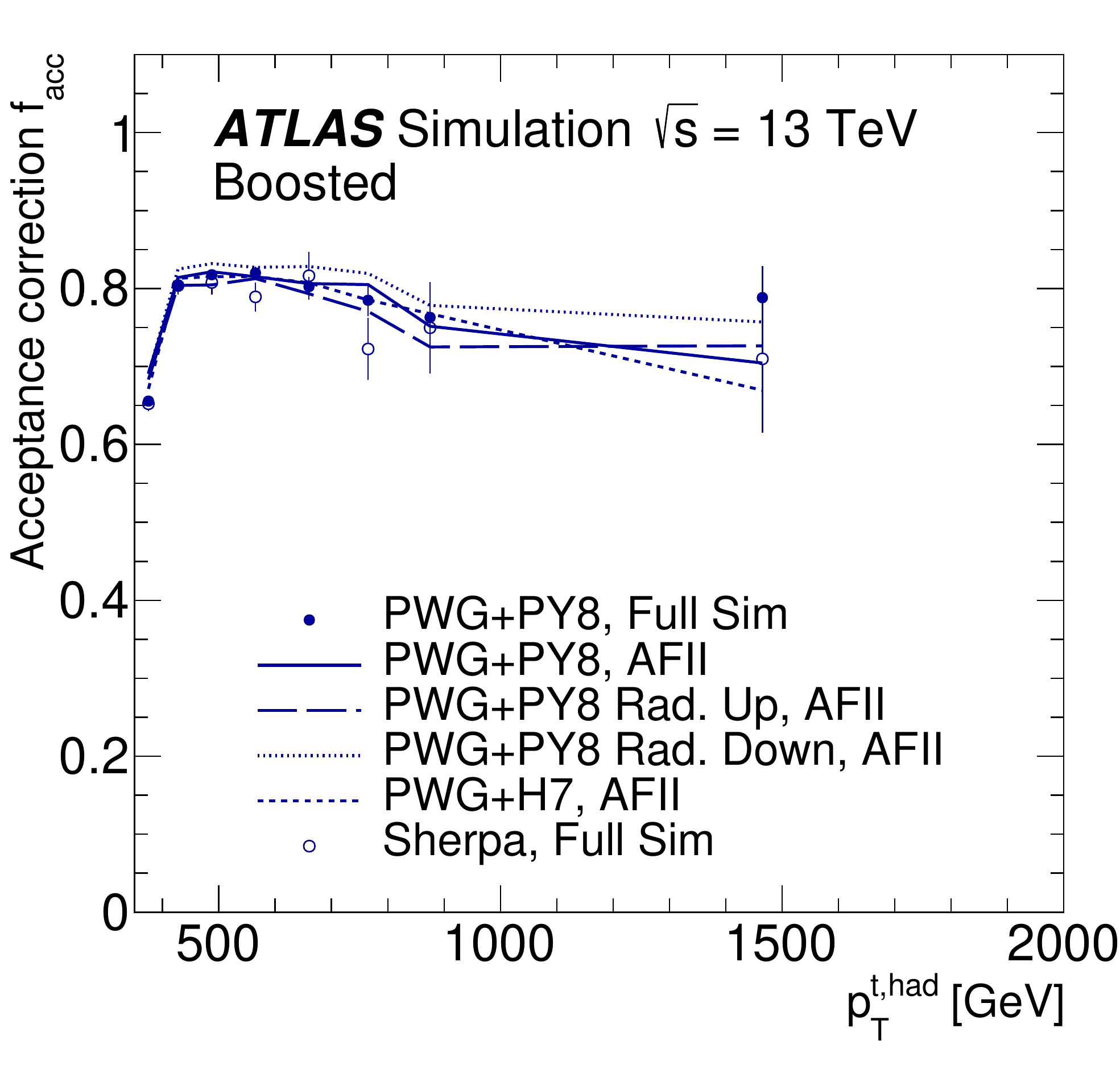}
\includegraphics[width=0.245\linewidth]{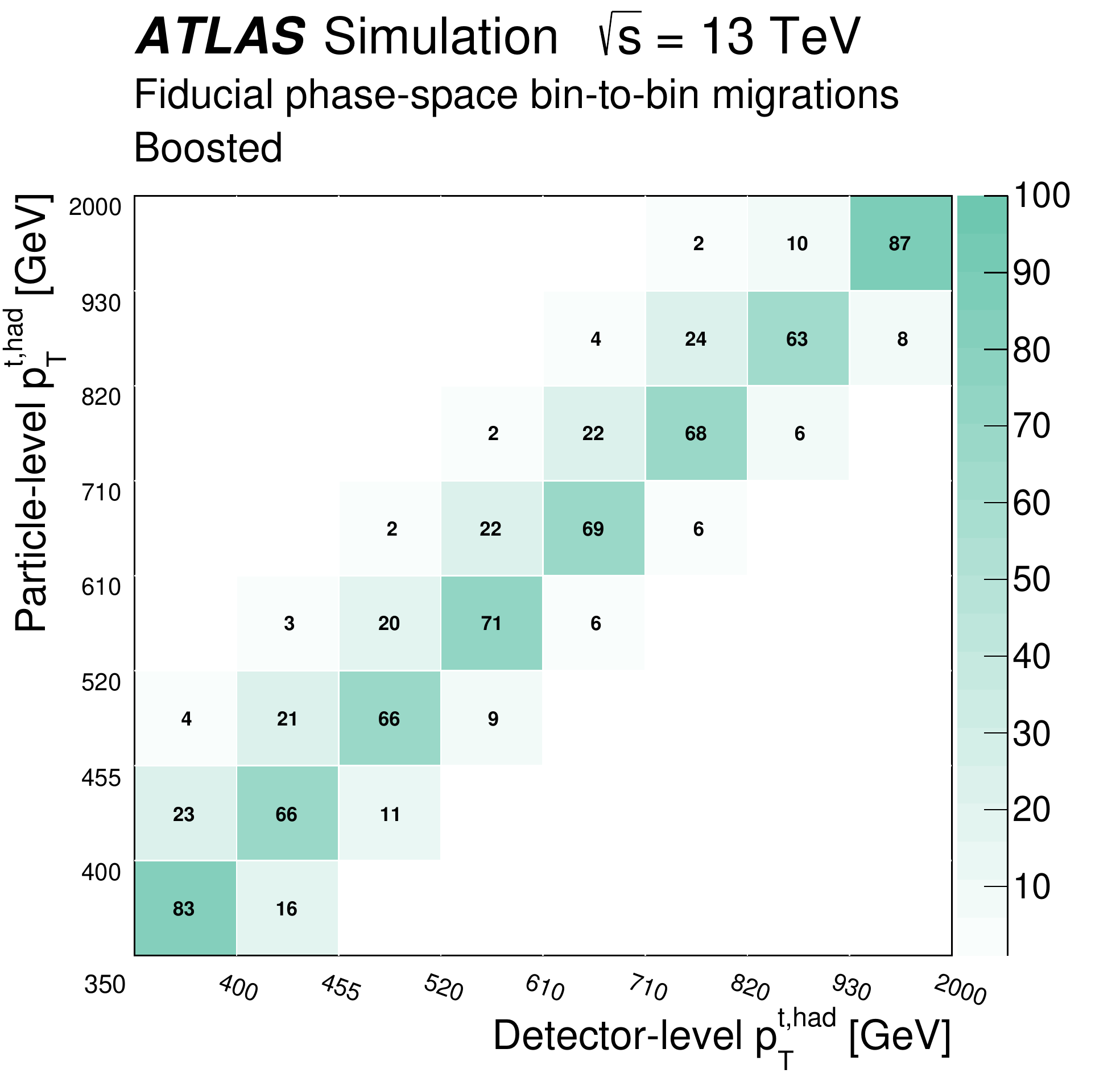}
\caption{Detector-level $p_\mathrm{T}$ (left) of the leading reclustered jet (hadronic top quark candidate); efficiency and dilepton corrections (middle); and the migration matrix (right) between the particle and detector levels~\cite{Aad:2019ntk}.}
\label{fig:reco_cmp}
\end{figure}

\begin{figure}
  \includegraphics[width=0.24\linewidth]{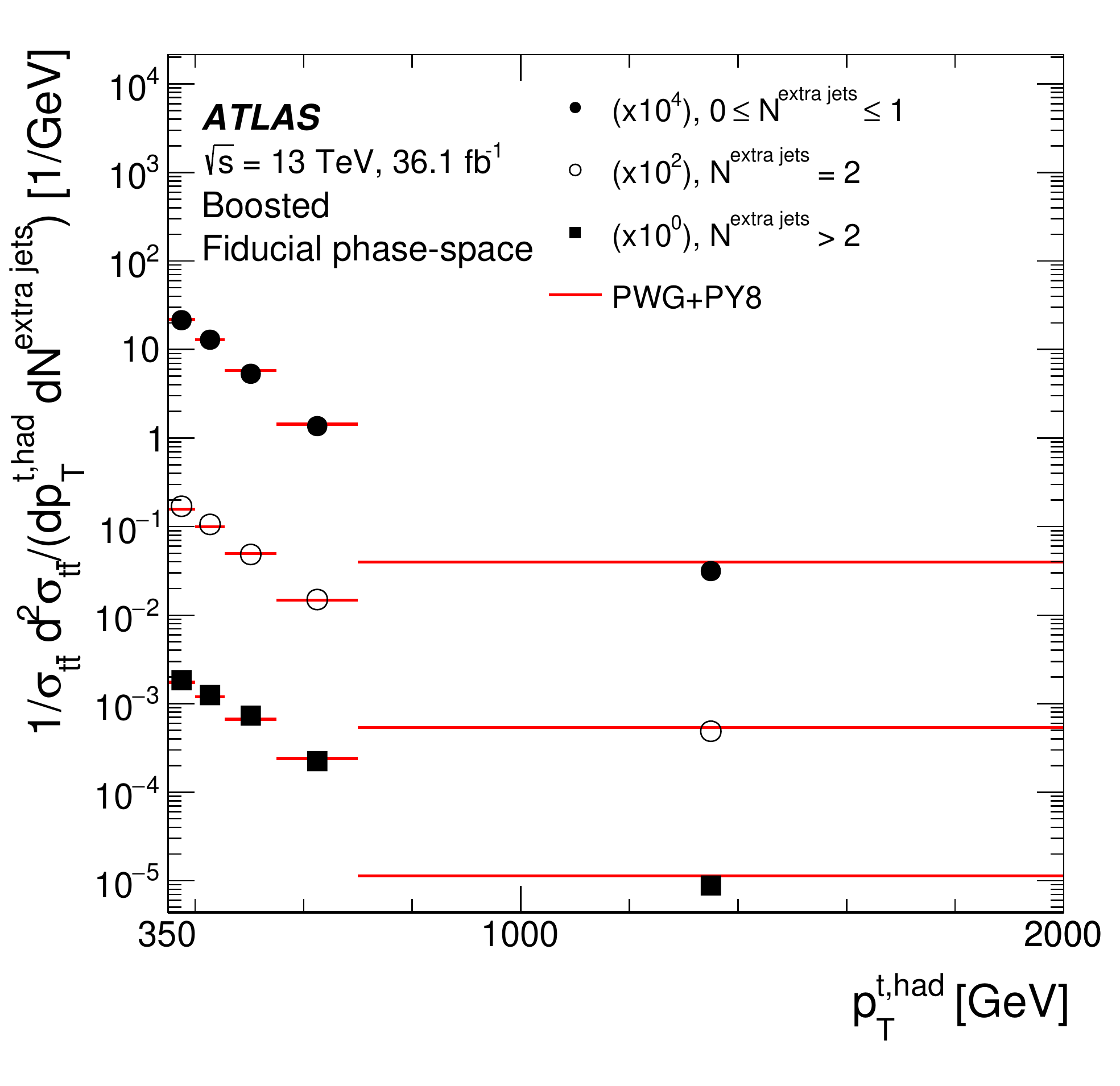}
  \includegraphics[width=0.70\linewidth]{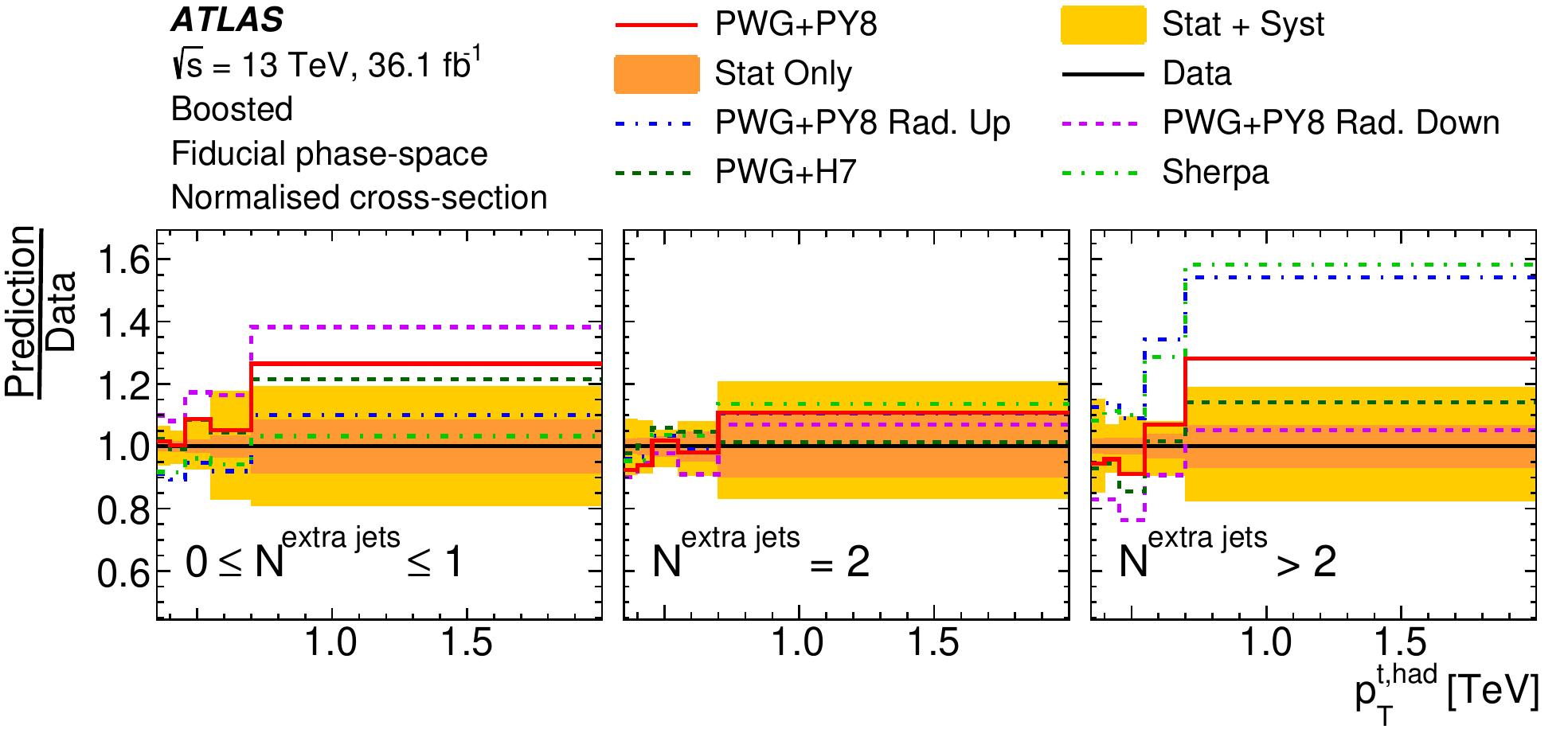}
\caption{Particle-level normalised differential cross-section as a function of the $p_\mathrm{T}$ of the hadronically decaying top quark in bins of the number of additional jets in the boosted topology compared with the prediction obtained with the \Powheg+\PythiaEight{} MC generator (left). Ratio of the measured cross-section to different Monte Carlo predictions (right). The bands represent the statistical and total uncertainty in the data~\cite{Aad:2019ntk}.}
\label{fig:reco_unf1}
\end{figure}

\begin{figure}[!h]
  \includegraphics[width=0.24\linewidth]{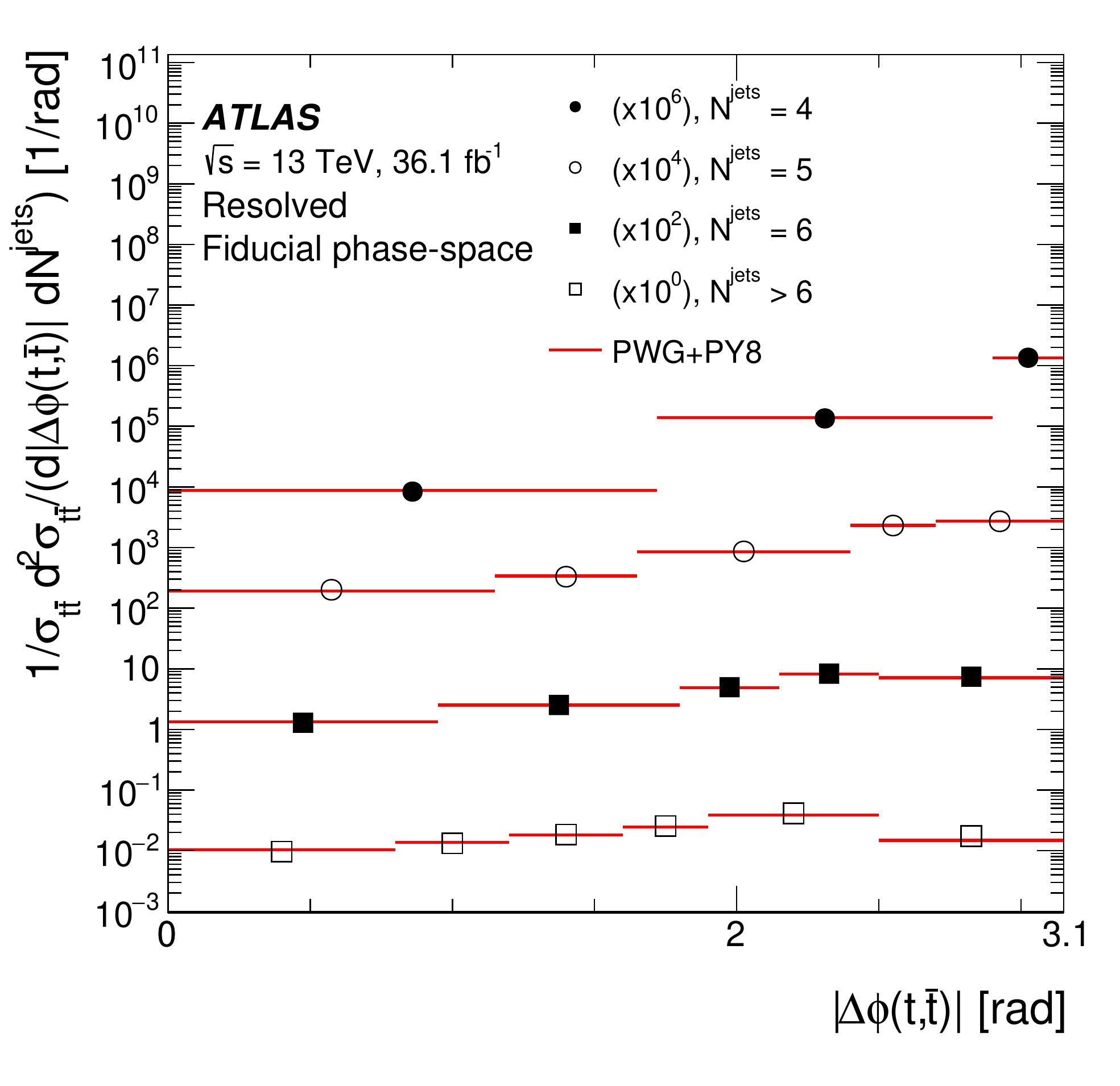}
  \includegraphics[width=0.70\linewidth]{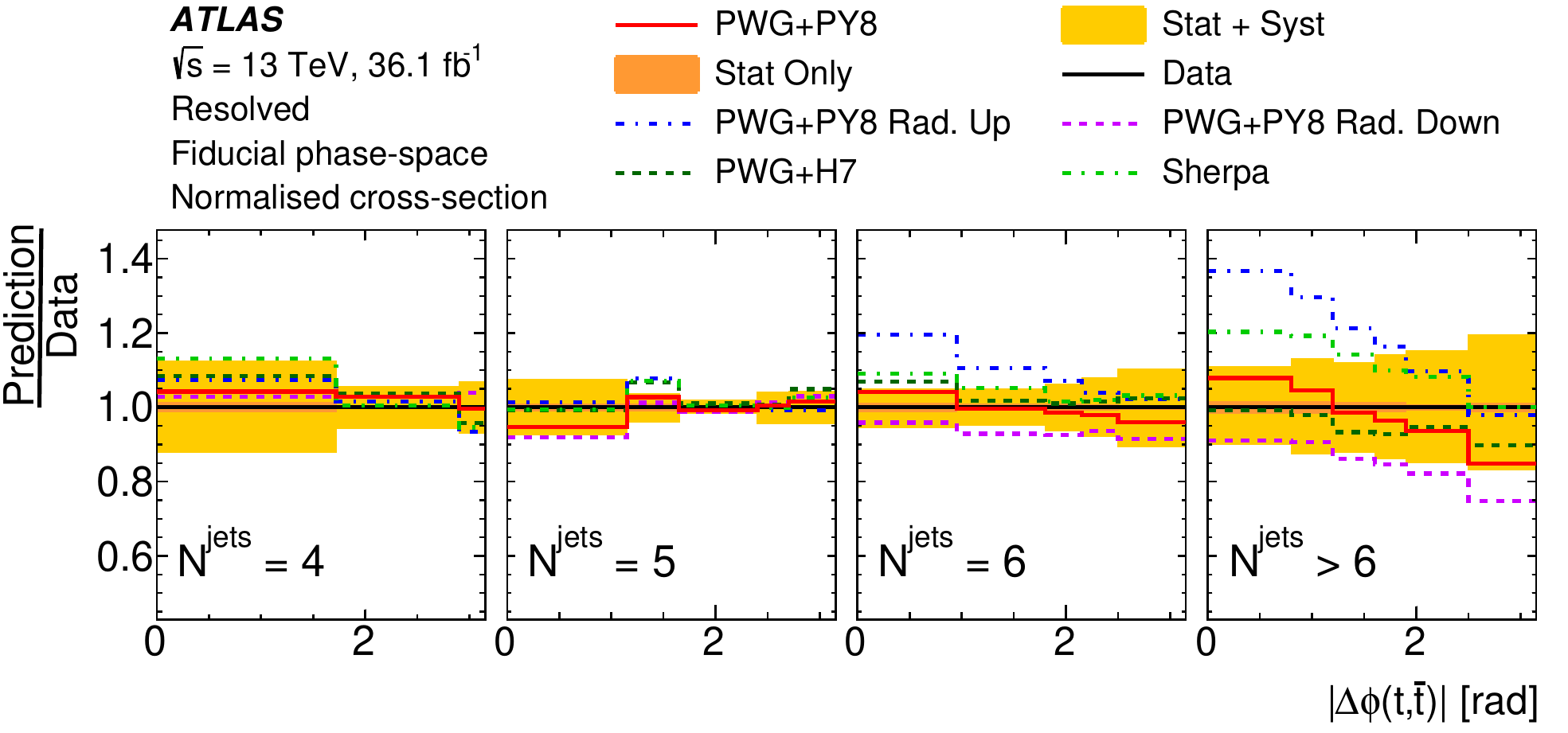}
  \caption{Particle-level normalised differential cross-section as a function of \deltaPhittbar{} in bins of the jet multiplicity in the resolved topology compared with the prediction obtained with the \Powheg+\PythiaEight{} MC generator (left). Ratio of the measured cross-section to  different Monte Carlo predictions (right). The bands represent the statistical and total uncertainty in the data~\cite{Aad:2019ntk}.}
  \label{fig:reco_unf2}
\end{figure}

Example of the detector-level data-to-prediction agreement, and unfolding ingredients for a 1D spectrum in the boosted topology is shown in~Figure~\ref{fig:reco_cmp}.
A choice of unfolded results to the fiducial particle-level phase space in the boosted and resolved topologies are shown in~Figures~\ref{fig:reco_unf1} and~\ref{fig:reco_unf2}.
For the boosted topology the 2-dimensional spectrum of the hadronic top-quark $p_\mathrm{T}$, in bins of additional jets, is shown in~Figure~\ref{fig:reco_unf1}. For the resolved topology the 2-dimensional spectrum of the angle between the two top quarks in the transverse plane, $\Delta\phi(t,\bar{t})$, in bins of the number of jets, is presented in~Figure~\ref{fig:reco_unf2}. The measurements can be used to tune the MC generators exploiting the data/prediction agreement.

\begin{figure}[!]
  \includegraphics[width=0.40\linewidth]{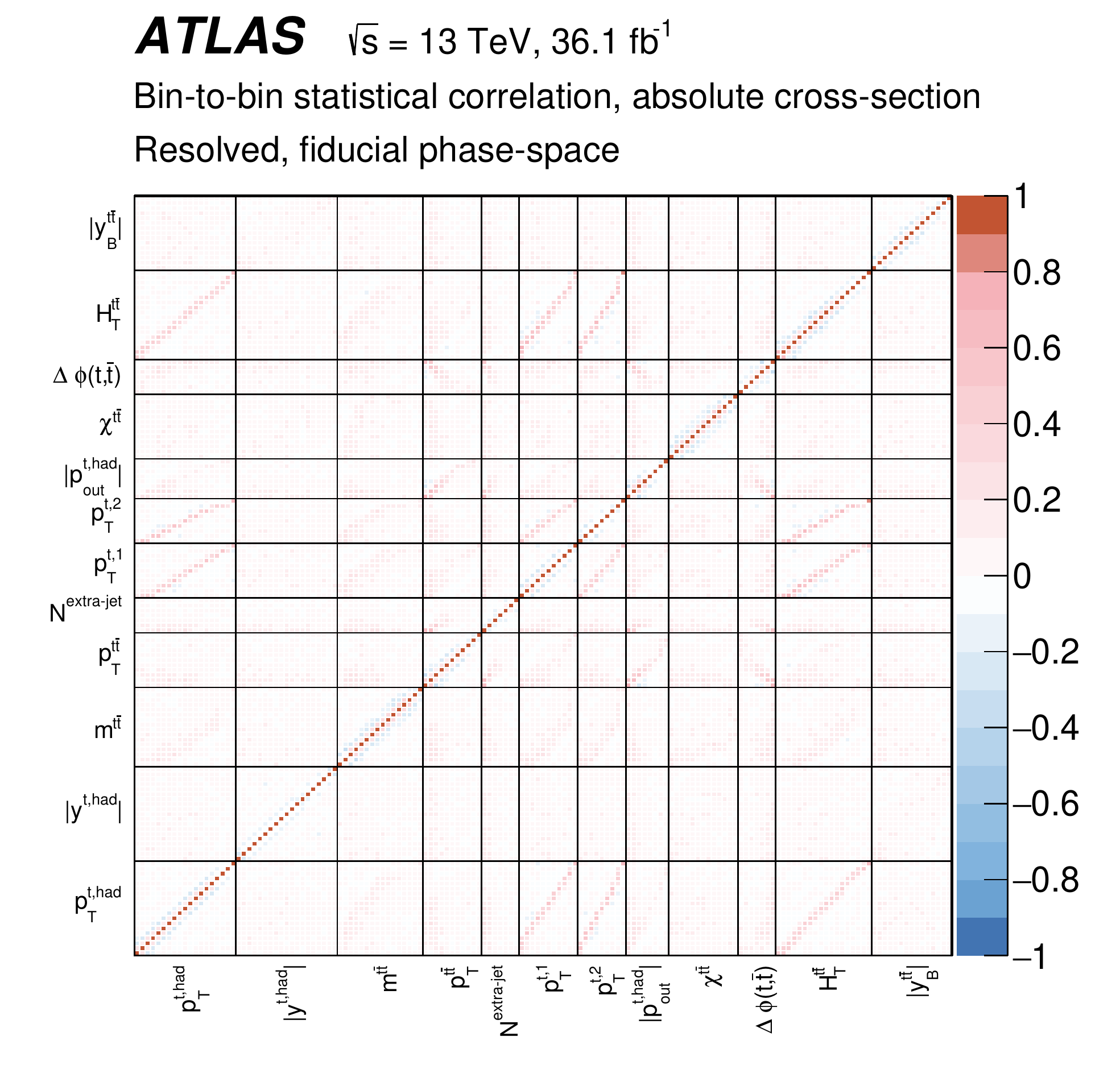}
  \includegraphics[width=0.58\linewidth]{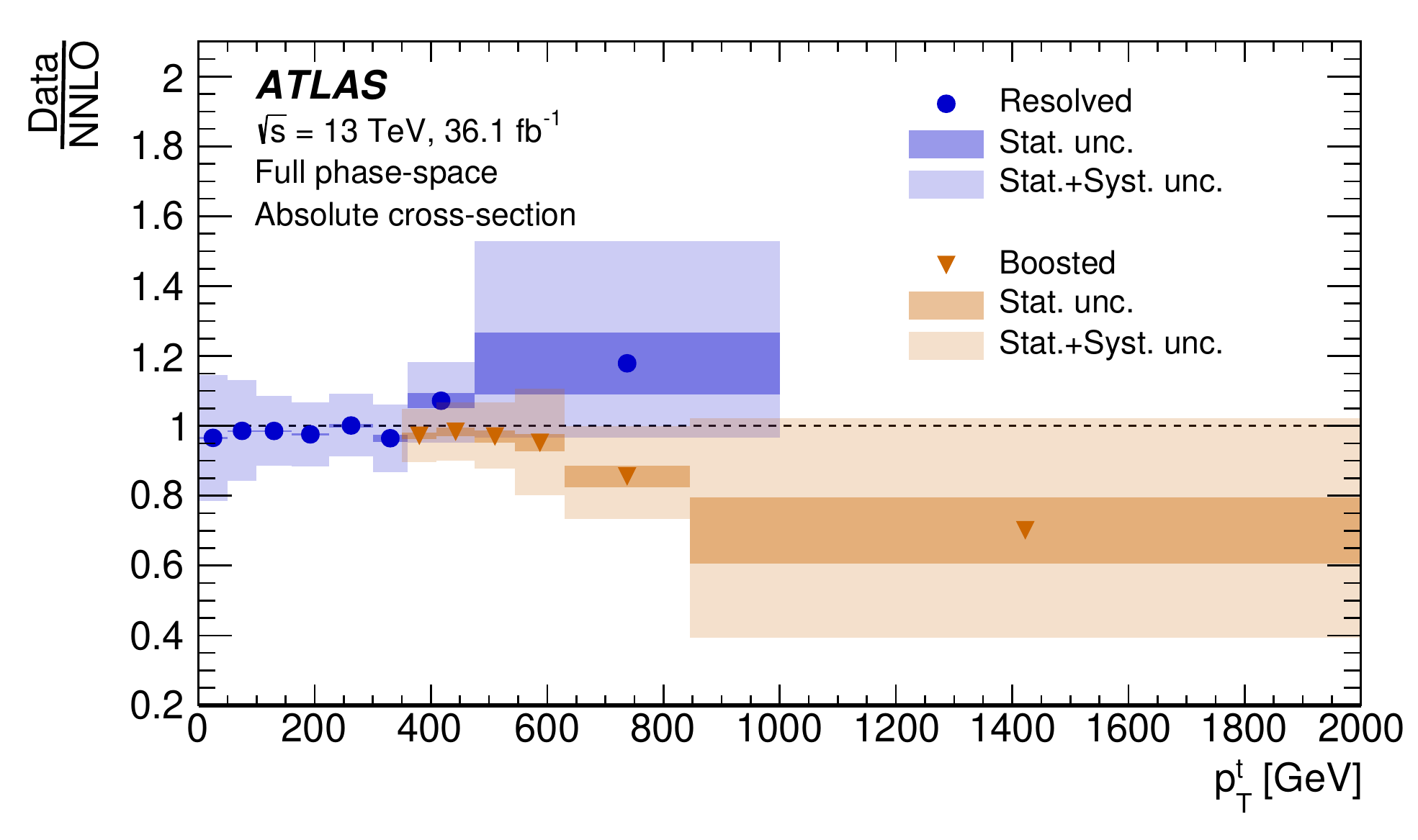}
    \caption{Correlation matrix between bins of distributions for simultaneous MC generators tuning (left).
      Comparison of the data ratio to NNLO QCD in both resolved and boosted regime at the parton level (right)~\cite{Aad:2019ntk}.}
    \label{fig:reco_corr}
 \end{figure}

A comparison of the resolved and boosted channels (although corresponding to different phase-spaces) is possible at the parton level in the form of a ratio to the NNLO QCD predictions~\cite{Czakon:2015owf,Czakon:2016dgf,Catani:2019hip}, as shown in~Figure~\ref{fig:reco_corr}. The same figure also includes an example of the statistical correlation matrix between bins of all the spectra in the resolved topology, enabling correlated MC generators tuning of several observables at the same time.

Agreement between the data and different initial and final state radiation MC settings are show in in~Table~\ref{tab:chi2} in terms of $p$-values for several 2D relative spectra in the resolved topology, showing the potential to tune the modelling parameters in order to diminish $\ttbar$ modelling uncertainties in other analyses where this process is a background.


\begin{table}[!h]
  \caption{The $p$-values from the comparison of the unfolded data to different settings of the \Powheg+\PythiaEight NLO MC generator (PP8)~\cite{Aad:2019ntk}.}
  \begin{center}
{\scriptsize
\begin{tabular}{|l | r @{/} l r  | r @{/} l r  | r @{/} l r|}
\hline 
Observable  & \multicolumn{3}{c|}{\textsc{PP8}} & \multicolumn{3}{c|}{\textsc{PP8} rad.~up} & \multicolumn{3}{c|}{\textsc{PP8} rad.~down} \\
  & \multicolumn{2}{c}{$\chi^{2}$/ndf} &  ~$p$-val& \multicolumn{2}{c}{$\chi^{2}$/ndf} &  ~$p$-val& \multicolumn{2}{c}{$\chi^{2}$/ndf} &  ~$p$-val \\ 
\hline 
\hline 
$H_{\mathrm{T}}^{t\bar{t}}\textrm{ vs }N^{\mathrm{extra \,\, jets}}$         & 9.7 & 19 & 0.96 &  57.9 & 19 & {\bf $<$0.01} &  19.4 & 19 & 0.43 \\
$|p_{\mathrm{out}}^{t,\mathrm{had}}|\textrm{ vs }N^{\mathrm{extra \,\, jets}}$ & 10.8 & 9 & 0.29 &  89.2 & 9 & {\bf $<$0.01} &  31.9 & 9 & {\bf $<$0.01} \\
$|\Delta\phi(t,\bar{t})| \textrm{ vs }N^{\mathrm{extra \,\, jets}}$     & 21.8 & 18 & 0.24 &  125.0 & 18 & {\bf $<$0.01} &  31.0 & 18 & 0.03 \\
$p_{\mathrm{T}}^{t,\mathrm{had}}\textrm{ vs }|p_{\mathrm{out}}^{t,\mathrm{had}}|$ & 10.5 & 12 & 0.57 &  74.5 & 12 & {\bf $<$0.01} &  25.3 & 12 & 0.01 \\
$p_{\mathrm{T}}^{t,\mathrm{had}}\textrm{ vs }N^{\mathrm{extra \,\, jets}}$ & 14.2 & 16 & 0.58 &  45.7 & 16 & {\bf $<$0.01} &  37.3 & 16 & {\bf $<$0.01} \\
$|y^{t\bar{t}}|\textrm{ vs }p_{\mathrm{T}}^{t\bar{t}}$ & 28.5 & 12 & {\bf $<$0.01} &  149.0 & 12 & {\bf $<$0.01} &  23.2 & 12 & 0.03 \\
\hline 
                             \end{tabular}
                           } 

   \label{tab:chi2}
   \end{center}
\end{table}

\section{Conclusions}
  New differential cross-section measurements for several observables in $\ttbar$ final states are presented, including the 2D spectra measured for the first time by ATLAS, in the $t\bar{t}\rightarrow \ell+$jets channel, with improved systematics uncertainties in the boosted regime owing to the usage of reclustered jets ($R=1$ jets built from $R=0.4$ anti-$k_t$ jets).
  The new measurements provide detailed information on the top-quark production and decay, allow precision tests of modern MC generators as well as latest fixed-order calculations. Good agreement is observed between predictions and the data within the reduced systematic uncertainties (which are typically dominant) w.r.t. previous ATLAS measurements. Comparisons to several MC setups and to different PDFs at the parton level have been performed~\cite{Aad:2019ntk}.


\clearpage
\Acknowledgements
The author would like to thank grant of MSMT, Czech Rep. ``CERN TRANSFER\_UP LTT17018'' for support.

\bibliographystyle{unsrt}
\bibliography{main}

\end{document}